\documentclass[conference]{IEEEtran}
\ifx\pdfoutput\undefined
\bibliography{mybib}
\usepackage{graphicx}
\else
\usepackage[pdftex]{graphicx}
\fi
\usepackage{url}
\usepackage{array}
\usepackage{stfloats}
\usepackage{amssymb}
\usepackage{amsmath}
\usepackage{amsthm}
\usepackage{cite}
\usepackage{enumerate}
\usepackage{epstopdf}
\begin{document}
 \title{ Constructive Interference through Symbol Level Precoding for  Multi-level Modulation}
 \author{
  Maha Alodeh \quad Symeon Chatzinotas \quad Bj\"{o}rn Ottersten\\
  \authorblockA{
  SnT-Interdisciplinary Centre for Security, Reliability and Trust, University of Luxembourg\\
  4, rue Alphonse Weicker, L-2721 Luxembourg\\
  e-mail:\{maha.alodeh, symeon.chatzinotas, bjorn.ottersten\}@uni.lu
               }
 } 
 
 \maketitle
 \begin{abstract}
The constructive interference concept in the downlink of multiple-antenna systems is addressed in this paper. The concept of the joint exploitation of the channel state information (CSI) and data information (DI) is discussed. Using symbol-level precoding, the interference between data streams
is transformed Under certain conditions into useful signal that can improve the signal to interference
noise ratio (SINR) of the downlink transmissions. In the previous work, different constructive interference precoding techniques have been proposed for the MPSK scenario. In this context, a novel constructive interference precoding technique that tackles the transmit power minimization (min-power) with
individual SINR constraints at each user's receivers is proposed assuming MQAM modulation. Extensive simulations are
performed to validate the proposed technique.
 
\let\thefootnote\relax\footnote{This work was supported by the National Research Fund (FNR) of
Luxembourg under the AFR grant (reference 4919957) for the project Smart Resource Allocation Techniques for Satellite Cognitive Radio. 
}
\end{abstract}
\section{Introduction}
Interference is one of the crucial factors that degrades the performance in wireless networks.
Exploiting the spatial dimension empowers the wireless system with additional dimension by adding multiple antennas at the communication terminals. In the literature, utilizing the time and frequency resources has been proposed to allow different users to share the resources without inducing harmful interference. The
concept of exploiting the users' spatial separation has been a fertile research domain for more than one
decade\cite{roy}. This can be implemented by adding multiple antennas at one or both
communication sides. Multiantenna transceivers provide the communication systems with more degrees
of freedom that can boost the performance if the multiuser interference is mitigated
properly. Exploiting the space dimension, to serve different users simultaneously
in the same time slot and the same frequency band through spatial division multiplexing (SDMA), has been investigated in \cite{roy}.

In this paper, the main idea is to constructively correlate
the interference among the spatial streams rather than fully decorrelate
them as in the conventional schemes \cite{haardt}. In \cite{Christos-1}, the interference in the scenario of BPSK and QPSK is classified into types: constructive and destructive. Based on this classification, a selective channel inversion scheme is proposed to eliminate the destructive interference while it preserves the constructive one to be received at the users' terminal. A more advanced scheme is proposed in \cite{Christos}, which  rotates the destructive interference to be received as useful signal with the constructive one. These schemes outperform the conventional precodings \cite{haardt} and show considerable gains. However, the anticipated
gains come at the expense of additional complexity at the system design level. Assuming
that
the channel coherence time is $\tau_{c}$, and the symbol period is $\tau_s$, with $\tau_c\gg\tau_s$ for slow fading
channels, the user precoder has to be recalculated with a frequency of $\frac{1}{\tau_c}$
in comparison with the symbol based precoder $\frac{1}{\min(\tau_c,\tau_s)}=\frac{1}{\tau_s}$. Therefore, faster precoder calculation and switching is needed in
the symbol-level precoding which can be translated to more expensive hardware.

In \cite{maha}-\cite{maha_TSP}, we have set the foundation for a symbol based precoding which opens new possibilities for exploiting the interference by establishing the connection between the constructive interference precoding and multicast. 
Moreover, several constructive interference precoding schemes have been proposed in \cite{maha_TSP}, including Maximum ratio transmission (MRT)-based algorithm  and objective-driven constructive interference techniques. The MRT based algorithm, titled as Constructive interference MRT (CIMRT), exploits the singular value decomposition (SVD) of the concatenated channel matrix. This enables the decoupled rotation using Givens rotation matrices  between the users' channels subspaces to ensure that the interference is received constructively at the users. On the other hand, the objective- driven optimization formulates the constructive interference problem by considering its relation to PHY-multicasting. However, all the previous contributions focus on utilizing the constructive interference assuming MPSK modulation,  
The contributions of this paper can be summarized in the following points:
\begin{itemize}
\item The previous works have discussed the constructive interference for M-PSK modulation. In this paper, we extend the constructive interference approach for MQAM modulation. The solution depends on the relation between the constructive interference precoding and PHY-layer multicasting.

\item Energy efficiency analysis is discussed to select the optimal SNR target for each modulation. Based on symbol error rate analysis and the power consumption, we find the SNR target that optimizes the energy efficiency.

\end{itemize}

\textbf{Notation}:  We use boldface upper and lower case letters for
 matrices and column vectors, respectively. $(\cdot)^H$, $(\cdot)^*$
 stand for Hermitian transpose and conjugate of $(\cdot)$. $\mathbb{E}(\cdot)$ and $\|\cdot\|$ denote the statistical expectation and the Euclidean norm, $\otimes$ denotes the kronecker
product, and $\mathbf{A}\succeq \mathbf{0}$ is used to indicate the positive
semidefinite matrix. $\angle(\cdot)$, $|\cdot|$ are the angle and magnitude  of $(\cdot)$ respectively. $\mathcal{R}(\cdot)$, $\mathcal{I}(\cdot)$
 are the real and the imaginary part of $(\cdot)$. 
\section{System and Signal Models}
\label{system}
We consider a single-cell multiple-antenna downlink scenario,
where a single BS is equipped with $M$
transmit antennas that serves $K$ user terminals,
each one of them equipped with a single receiving antenna. The adopted
modulation technique is M-QAM.
We assume a quasi static block fading channel $\mathbf{h}_j\in\mathbb{C}^{1\times
M}$ between
the BS antennas and the $j^{th}$ user, where the received signal at
j$^{th}$ user is
written as
\begin{eqnarray}
y_j[n]&=&\mathbf{h}_j\mathbf{x}[n]+z_j[n].
\end{eqnarray} $\mathbf{x}[n]\in\mathbb{C}^{M\times 1}$ is the transmitted symbol sampled signal vector at time $n$ from the multiple antennas
transmitter and  $z_j$ denotes the noise at $j^{th}$ receiver, which is assumed i.d.d  complex Gaussian distributed variable $\mathcal{CN}(0,1)$. A compact formulation
of the received signal at all users' receivers can be written as
\vspace{-0.2cm}
\begin{eqnarray}
\mathbf{y}[n]&=&\mathbf{H}\mathbf{x}[n]+\mathbf{z}[n].
\end{eqnarray}
Let $\mathbf{x}[n]$ be written as $\mathbf{x}[n]=\sum^K_{j=1}\sqrt{p_j[n]}\mathbf{w}_j[n]d_j[n]$,
where $\mathbf{w}_j$ is the $\mathbb{C}^{M\times
1}$ unit power precoding vector for the user $j$. The received signal at $j^{th}$
user ${y}_j$ in $n^{th}$ symbol period is given by
\begin{eqnarray}
\label{rx_o}
\hspace{-1cm}{y}_j[n]=\sqrt{p_j[n]}\mathbf{h}_j\mathbf{w}_j[n] d_j[n]+\displaystyle\sum_{k\neq j}\sqrt{p_k[n]}\mathbf{h}_j\mathbf{w}_k[n]
d_k[n]+z_j[n]
\end{eqnarray}
where $p_j$ is the allocated power to the $j^{th}$ user.
Notice that the transmitted signal $\mathbf{d}\in\mathbb{C}^{K\times 1}$
includes the uncorrelated data symbols $d_k$ for all users with $\mathbb{E}[{|d_k|^2}] = 1$.
It should be noted that both CSI and data information (DI) are available at the transmitter side. From now on, we assume that the precoding design is performed at each symbol period and accordingly we drop the time index for the sake of notation.
\section{Conventional Multiuser Precoding Techniques}
\label{traditional}
The main goal of transmit beamforming is to increase the signal power at
the intended user and mitigate the interference to non-intended users. This can be mathematically translated to a design problem that targets
 beamforming vectors to have maximal inner products with 
 the intended channels and minimal inner products with the non-intended ones.
Several approaches have been
proposed including minimizing the sum power while satisfying
 a set of SINR constraints\cite{mats} and  maximizing the jointly achievable SINR
margin under a  power constraint\cite{boche}. 
In any scenario,  
the generic received signal can be formulated as 
\begin{eqnarray}\nonumber
\label{interference}
\hspace{-0.7cm}\mathbf{y}[n]&=&\mathbf{H}\mathbf{x}[n]+\mathbf{z}[n]=\mathbf{H}\mathbf{W}[n]\mathbf{P}^{\frac{1}{2}}[n]\mathbf{d}[n]+\mathbf{z}[n]\\
&=&\begin{bmatrix}\underset{\text{desired}}{\underbrace{a_{11}}}&\underset{\text{interference}}{\underbrace{a_{12}}}&\hdots&\underset{\text{interference}}{\underbrace{a_{1K}}}\\
a_{21}&\underset{\text{desired}}{\underbrace{a_{22}}}&\dots&a_{2K}\\
\vdots&\vdots&\vdots&\vdots\\
\underset{\text{interference}}{\underbrace{a_{K1}}}&\underset{\text{interference}}{\underbrace{a_{K2}}}&\dots&\underset{\text{desired}}{\underbrace{a_{KK}}}
\end{bmatrix}\begin{bmatrix}d_1\\
\vdots\\
d_K\end{bmatrix}+\mathbf{z}.
\end{eqnarray}
The corresponding SINR of user $j$ can be expressed as\\
\begin{eqnarray}
\gamma_{j}=\frac{p_j\|\mathbf{h}_j\mathbf{w}_j\|^2}{\sum^K_{i=1,i\neq j}p_i\|\mathbf{h}_j\mathbf{w}_i\|^2+\sigma^2}=\frac{|a_{jj}|^2}{\sum^K_{i=1,i\neq
j}|a_{ji}|^2+\sigma^2}.
\end{eqnarray}

Symbol-level precoding tries to go beyond this conventional look at the interference.
This precoding can under certain conditions convert the inner product 
with the non-intended channels into useful power by maximizing them but
with the specific directions to which constructively add-up at each user receivers.
Taking into account  the I/Q plane of the symbol detection, the constructive interference is achieved by using the interfering signal vector to move the received point deeper into the correct detection region. Considering
that each user receives a constructive interference from other users' streams,
the received signal can be written as
\vspace{-0.1cm}
\begin{eqnarray}
y_j[n]=\sum^K_{i=1}\underset{a_{ji}[n]d_{j}[n]}{\underbrace{\sqrt{p_j[n]}\mathbf{h}_j\mathbf{w}_i[n]d_i[n]}}+z_j[n].
\end{eqnarray}
 This yields the SINR expression for M-PSK symbols as 
\begin{eqnarray}
\gamma_{j}[n]=\frac{\|\sum^K_{i=1}\sqrt{p_j[n]}\mathbf{h}_j\mathbf{w}_i[n]\|^2}{\sigma^2}=\frac{|\sum^K_{i=1}a_{ji}|^2}{\sigma^2}.
\end{eqnarray}
Different precoding techniques that redesign the terms $a_{ji},j\neq i$ to constructively
correlate them with $a_{jj}$ are proposed in the next sections (\ref{powmin}).
\subsection{Power constraints for user based and symbol based precodings}
In the conventional user based precoding, the transmitter needs to precode every $\tau_{c}$
which means that the power constraint has to be satisfied along the coherence time
$\mathbb{E}_{\tau_c}\{\|\mathbf{x}\|^2\}\leq
P$. Taking the expectation of $\mathbb{E}_{\tau_c}\{\|\mathbf{x}\|^2\}=\mathbb{E}_{\tau_c}\{tr(\mathbf{W}\mathbf{d}\mathbf{d}^H\mathbf{W}^H)\}$,
and since $\mathbf{W}$ is fixed along $\tau_c$, the previous expression can
be reformulated as $tr(\mathbf{W}\mathbb{E}_{\tau_c}\{\mathbf{d}\mathbf{d}^H\}\mathbf{W}^H)=tr(\mathbf{W}\mathbf{W}^H)=\sum^K_{j=1}\|\mathbf{w}_j\|^2$,
where $\mathbb{E}_{\tau_c}\{\mathbf{d}\mathbf{d}^H\}=\mathbf{I}$ due to uncorrelated
symbols over $\tau_c$.
However, in symbol level precoding the power constraint should be guaranteed
for each symbol vector transmission namely for each $\tau_s$. In this case
the power constraint equals to $\|\mathbf{x}\|^2=\mathbf{W}\mathbf{d}\mathbf{d}^H\mathbf{W}^H=\|\sum^K_{j=1}\mathbf{w}_jd_j\|^2$.
In the next sections, we characterize the constructive interference and show
how to exploit it in the multiuser downlink transmissions\footnote{From now on, we assume that the transmission changes at each symbol and we drop the time index for the ease of notation }.
\section{Constructive interference for power minimization}
 \label{powmin}
 The interference among the simultaneous spatial streams
leads to deviation of the received symbols from their detection region. However, this interference can be designed to push the received symbols further into the correct detection region assuming MPSK modulation and, as a consequence it enhances the system performance \cite{maha}-\cite{maha_crowncom}. However the case is different for MQAM, the constructive interference can be exploited to push the outer constellation symbols deeper in their detection regions. For the inner constellation symbols, this cannot be applied directly. Assuming both DI and CSI are available at the transmitter, the 
 cross correlation between the $k^{th}$ data stream and the $j^{th}$ user can be formulated as:
\begin{equation}
\rho_{jk}=\frac{\mathbf{h}_{j}\mathbf{h}^H_k}{\|\mathbf{h}_{j}\|\|\mathbf{h}_k\|}.
\end{equation}

 \subsection{Constructive Interference Power Minimization Precoding for MQAM modulation (MCIPM)}

Based on the definition of constructive interference, we should design the constructive interference precoders by guaranteeing that the sum of the
precoders and data symbols pushes the received signal deeper in the correct detection region for outer constellation symbols and achieves the exact symbols for the inner constellation ones. 
Therefore, the optimization that
 minimizes the transmit power and grants
 the constructive reception of the transmitted data symbols can be written
 as 
\vspace{-0.2cm}
\begin{eqnarray}
\label{powccm}
&\hspace{-0.1cm}\mathbf{w}_k(d_j,\mathbf{H},\boldsymbol\zeta)
&\hspace{-0.05cm}=\arg\underset{\mathbf{w}_1,\hdots,\mathbf{w}_K}{\min}\quad \|\sum^K_{k=1}\mathbf{w}_kd_k\|^2\\\nonumber
&s.t.&\mathcal{C}_1,\mathcal{C}_2
\end{eqnarray}

For the received signal at $j^{th}$ user, we denote $\alpha^r_j$,$\alpha^i_j$ as the in-phase and the quadrature components respectively. $\alpha^r_j$, $\alpha^i_j$ can be mathematically formulated as
\begin{eqnarray}
\alpha^r_j=\frac{\mathbf{h}_j\sum_{j}\mathbf{w_j}d_j+(\mathbf{h}_j\sum_j\mathbf{w}_jd_j)^H}{2}\\
\alpha^i_j=\frac{\mathbf{h}_j\sum_{j}\mathbf{w_j}d_j-(\mathbf{h}_j\sum_j\mathbf{w}_jd_j)^H}{2i}
\end{eqnarray}

$\mathcal{C}_1$, $\mathcal{C}_2$  can be formulated to guarantee that the received signal lies in the correct detection region, which depends on the data symbols. A detailed formulation for $\mathcal{C}_1$, $\mathcal{C}_2$ can be expressed as

\begin{itemize}
\item For the inner-constellation symbols, the constraints $\mathcal{C}_1$, $\mathcal{C}_2$ should guarantee that the received signals achieve the exact constellation point. For 16-QAM as depicted in Fig. (\ref{qam}), the symbols marked by 1 should be received with the exact symbols. The constraints can be written as    
\begin{eqnarray}\nonumber
&\mathcal{C}_1:&\alpha^r_{j}=\sqrt{\zeta_j}\mathcal{R}\{d_j\}\\\nonumber
&\mathcal{C}_2:&\alpha^i_{j}=\sqrt{\zeta_j}\mathcal{I}\{d_j\}
\end{eqnarray}

\item Outer constellation symbols, the constraints $\mathcal{C}_1$, $\mathcal{C}_2$ should guarantee the received signals lie in the correct detection. For 16-QAM as depicted in Fig. (\ref{qam}), the symbols marked by 2 should be received with the exact symbols. The constraints can be written as     
\begin{eqnarray}\nonumber
&\mathcal{C}_1:&\alpha^r_{j}\gtreqless\sqrt{\zeta_j}\mathcal{R}\{d_j\}\\\nonumber
&\mathcal{C}_2:&\alpha^i_{j}=\sqrt{\zeta_j}\mathcal{I}\{d_j\}
\end{eqnarray}
\begin{eqnarray}\nonumber
&\mathcal{C}_1:&\alpha^r_{j}=\sqrt{\zeta_j}\mathcal{R}\{d_j\}\\\nonumber
&\mathcal{C}_2:&\alpha^i_{j}\gtreqless\sqrt{\zeta_j}\mathcal{I}\{d_j\},
\end{eqnarray}
\item Outermost constellation symbols, the constraints $\mathcal{C}_1$, $\mathcal{C}_2$ should guarantee the received signals lie in the correct detection. For 16-QAM as depicted in Fig. (\ref{qam}), the symbols marked by 3 should be received with the exact symbols. The constraints can be written as     
\begin{eqnarray}\nonumber
&\mathcal{C}_1:&\alpha^r_{j}\gtreqless\sqrt{\zeta_j}\mathcal{R}\{d_j\}\\\nonumber
&\mathcal{C}_2:&\alpha^i_{j}\gtreqless\sqrt{\zeta_j}\mathcal{I}\{d_j\}.
\end{eqnarray} 
The sign $\gtreqless$ indicates that the symbols should locate in the correct detection region, for the symbols in the first quadrant $\gtreqless$ mean $\geq$.
\end{itemize}
where $\zeta_j$ is the SNR target for the $j^{th}$ user that should
 be granted by the transmitter, and ${\boldsymbol\zeta}=[\zeta_1,\hdots,\zeta_K]$ is the vector that contains all the SNR targets that should be guaranteed by BS to each user. This way, each receiver can correctly scale its margins during the symbol detection. The set of constraints $\mathcal{C}_1$ $\mathcal{C}_2$ 
 guarantees that each user receives its corresponding data symbol $d_j$ correctly. If we assume that all users' data symbols lie at the outer constellation points at certain instant and if we denote $\mathbf{x}=\sum^K_{k=1}\mathbf{w}_kd_k$,
 a formulation of (\ref{powccm}) can be expressed as 
\vspace{-0.1cm} 
\begin{eqnarray}
\label{powa}
\mathbf{x}=\hspace{-0.9cm}&\quad&\arg\underset{\mathbf{x}}{\min}\quad \|\mathbf{x}\|^2\\\nonumber
&s.t.&\begin{cases}\mathcal{C}1:\frac{\mathbf{h}_j\mathbf{x}+(\mathbf{h}_j\mathbf{x})^H}{2}{\geq}\sigma\sqrt{\zeta_j}\mathcal{R}\{d_j\},\forall
j\in K\\
\mathcal{C}2:\frac{\mathbf{h}_j\mathbf{x}-(\mathbf{h}_j\mathbf{x})^H}{2i}{\geq}\sigma\sqrt{\zeta_j}\mathcal{I}\{d_j\},
\forall
j\in K.
\end{cases}
\end{eqnarray}
  The solution for (\ref{powa}) can be found by writing    
the Lagrangian function  as follows
\begin{eqnarray}\nonumber
\label{lagrange}
\hspace{-1.9cm}&\mathcal{L}&(\mathbf{x})=\|\mathbf{x}\|^2\\\nonumber
&+&\sum_j{\mu_j}\bigg(-0.5i\frac{\small(\mathbf{h}_j\mathbf{x}-\mathbf{x}^H\mathbf{h}^H_j\small)}{\sigma\sqrt{\zeta_{j}}}-\mathcal{I}\{d_j\}\bigg)\\
&+&\sum_j{\alpha_j}\bigg(0.5\frac{\small(\mathbf{h}_j\mathbf{x}+\mathbf{x}^H\mathbf{h}^H_j\small)}{\sigma\sqrt{\zeta_{j}}}
-\mathcal{R}\{d_j\}\bigg)
\end{eqnarray}
where $\mu_j$ and $\alpha_j$ are the Lagrangian dual variables. It should be noted that the Lagrange function is dependent on the set of constraints related to the symbols. For example, the Lagrange function changes with set the set of the data that should be sent to each user. The derivative for the Lagrangian function can be written as
\vspace{-0.1cm}
\begin{eqnarray}
\frac{d\mathcal{L}(\mathbf{x})}{d\mathbf{x}^*}=\mathbf{x}+0.5i\sum_j\mu_j\mathbf{h}^H_j
+0.5\sum_j\alpha_j\mathbf{h}^H_j.
\end{eqnarray}
\vspace{-0.15cm}
By equating this term to zero, $\mathbf{x}$ can be written as
\hspace{-0.2cm}\begin{eqnarray}\nonumber
\label{CIPM}
\hspace{-0.5cm}\mathbf{x}&=&i\sum^K_{j=1}\mu_j\mathbf{h}^H_j
+0.5\sum_j\alpha_j\mathbf{h}^H_j\\
 &=&\sum^K_{j=1}\nu_j\mathbf{h}^H_j,\forall
j\in K
\end{eqnarray}
where $\nu_j\in \mathbb{C}$. The optimal values of the Lagrangian variables $\mu_j$ and $\alpha_j$ can
be found by substituting $\mathbf{w}$ in the constraints (\ref{powa}) which result in solving the set of $2K$ equations (\ref{setoo}).
 The final 
precoder can be found by substituting all $\mu_j$ and $\alpha_j$ in (\ref{CIPM}). A generic solution for any set of simultaneous data symbols can be formulated as
\vspace{-0.2cm}
\begin{eqnarray}
\mathbf{x}=\sum^K_{j=1}\nu_j\mathbf{h}^H_j.
\end{eqnarray}
It can be noted that the precoding is a summation of maximum ratio transmissions precoding for all users.
\begin{figure*}[t]
\begin{tabular}[t]{c}
\begin{minipage}{17 cm}
 \begin{eqnarray}
\label{multicasteq}
\begin{array}{cccc}
\label{setoo}
0.5K\|\mathbf{h}_1\|(\sum_k(-\mu_k+\alpha_ki)\|\mathbf{h}_k\|\rho_{1k}&-&\sum_k(-\mu_k+\alpha_ki)\|\mathbf{h}_k\|\rho^{*}_{1k}){\geq}\sigma\sqrt{\zeta^{}_{1}}\mathcal{I}(d_1)\\
0.5K\|\mathbf{h}_1\|(\sum_k(-\mu_ki-\alpha_k)\|\mathbf{h}_k\|\rho_{1k}&+&\sum_k(-\mu_ki-\alpha_k)\|\mathbf{h}_k\|\rho^{*}_{1k}){\geq}\sigma\sqrt{\zeta^{}_{1}}\mathcal{R}(d_1)\\
\quad&\vdots&\\
0.5K\|\mathbf{h}_K\|(\sum_k(-\mu_k+\alpha_ki)\|\mathbf{h}_k\|\rho_{Kk}&-&\sum_k(-\mu_k+\alpha_ki)\|\mathbf{h}_k\|\rho^{*}_{Kk}){\geq}\sigma\sqrt{\zeta_{K}}\mathcal{I}(d_K)\\
0.5K\|\mathbf{h}_K\|(\sum_k(-\mu_ki-\alpha_k)\|\mathbf{h}_k\|\rho_{Kk}&+&\sum_k(-\mu_ki-\alpha_k)\|\mathbf{h}_k\|\rho^{*}_{Kk}){\geq}\sigma\sqrt{\zeta_{K}}\mathcal{R}(d_K)\\
\end{array}
\end{eqnarray}
\end{minipage}\\
\hline
\hline
\end{tabular}
\end{figure*}
\begin{figure}[t]
\hspace{-0.9cm}\includegraphics[scale=0.38]{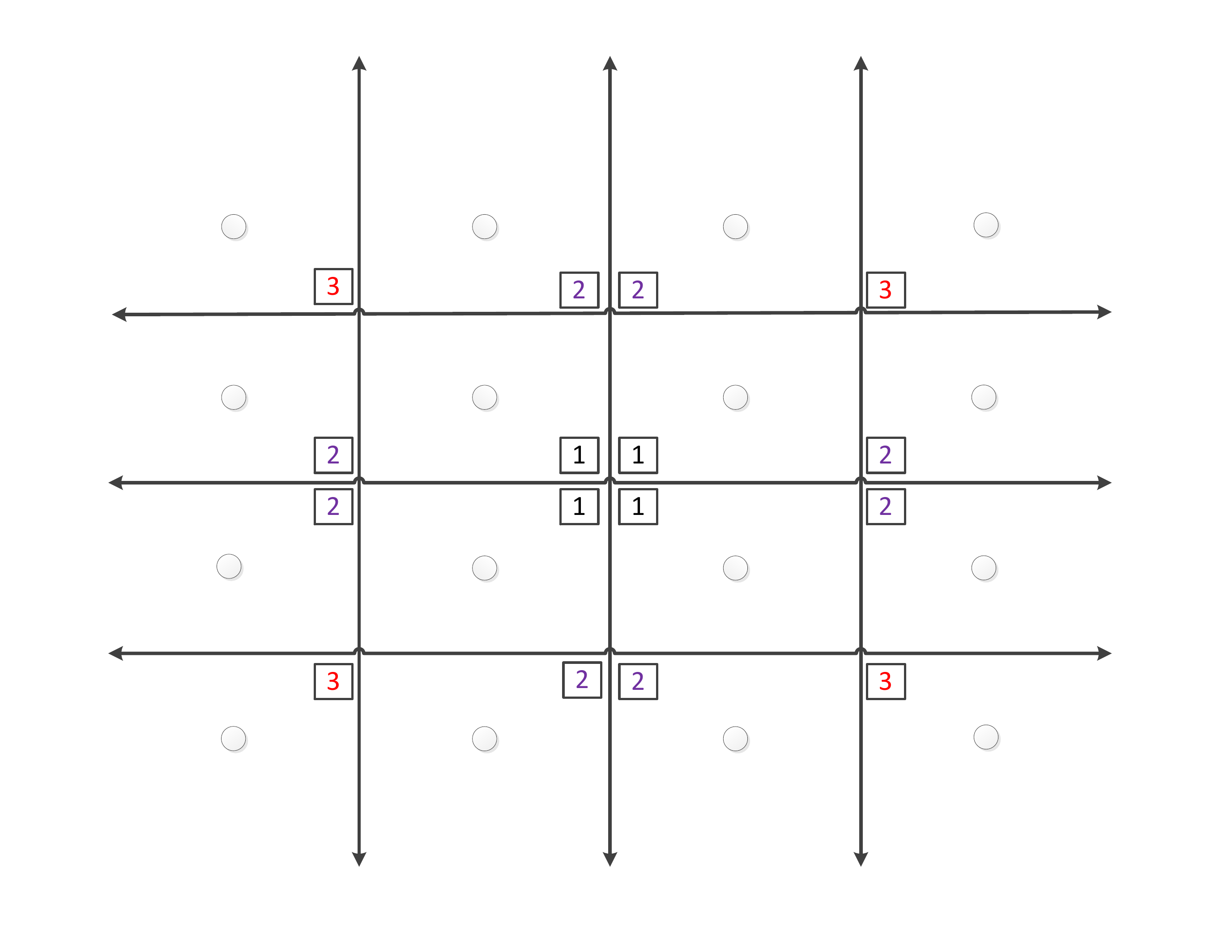}
\vspace{-0.7cm}\caption{\label{qam}\textit{\label{qam}\small 16-QAM modulation}}
\end{figure}

\section{Energy efficiency analysis}

Due to the noise at the receiver, the detected symbols can deviate from the correct detection region. The effective rate $\bar{R_j}$ (i.e. goodput) for $j^{th}$ user can be expressed as  
\begin{eqnarray}
\label{effective_r}
\bar{R}_j\approx R_j\times \big(1-SER_j (\zeta_j,z_j)\big)
\end{eqnarray}
$R_{j}$ is the $j^{th}$  user target rate of the employed modulation and $SER_j$ denotes the symbol error rate of the $j^{th}$ users. 
From (\ref{effective_r}), it can be noticed that increasing the SNR targets $\zeta_j$  reduces the probability of errors resulted from the noise, and as a result it enhances the effective rate. 
\subsection{Energy efficiency analysis}
Increasing the SNR target reduces the SER while it increases the power consumption to achieve the SNR target. To find the optimal balance between these two aspects, the system energy efficiency metric is proposed to find how many bits can be conveyed correctly to the receivers per energy unit. The system energy efficiency can be defined as 
\begin{eqnarray}
\label{eta}
\eta(\boldsymbol\zeta)= \frac{\sum^K_{j=1} \bar{R}_j\Big(SER_j(\zeta_j)\Big)}{P(\boldsymbol\zeta)}
\end{eqnarray}
where $P(\boldsymbol \zeta)=\|\mathbf{x}(\mathbf{H},\mathbf{d},\boldsymbol\zeta)\|^2$. 
It should be noted that the energy efficiency is a function of the SNR target $\zeta_j$ since it increases the transmit power amount required to achieve the target rate. Changing the SNR target affects both the numerator and the denominator in (\ref{eta}) by increasing the effective rate and transmit power respectively.  
\section{Numerical Results}

The channel between the base station
and $j^{th}$ user terminal is characterized by
$\mathbf{h}_j=\sqrt{\gamma_{\circ}}\mathbf{h}^{'}_j$
where $\mathbf{h}^{'}_j\sim\mathcal{CN}(0,\mathbf{1})$, and $\gamma_{\circ}$ is
the average channel power. For the sake of comparison, we plot the performance the physical layer multicasting as a bound\cite{multicast} 
 \begin{eqnarray}
 \mathbf{Q}=\arg\underset{\mathbf{Q}}{\min}\quad \text{trace}(\mathbf{Q}), s.t.\quad \mathbf{h}_j\mathbf{Q}\mathbf{h}^H_j\geq \zeta_{j},\forall j\in K.
 \end{eqnarray}
 
It should be noted for the sake of comparison between 4-QAM, 8-QAM, and 16QAM, the constellations are scaled mathematically to have average power of all constellation symbols
 should equal to 1. The scaling factor equals to $1$, $\frac{1}{\sqrt{3}}$ and $\frac{1}{\sqrt{5}}$ for 4-QAM, 8-QAM and 16-QAM respectively.
 
 For 8-QAM, the constraints $\mathcal{C}_1$, $\mathcal{C}_2$ for each symbol can be written in details as
 
 \begin{eqnarray}\nonumber
\mathcal{C}_1=\begin{cases}\alpha_r=\sigma\sqrt{\frac{\zeta_j}{3}}\mathcal{R}\{d_j\}, d_j=\frac{\pm 1\pm i}{\sqrt{2}}\\
\alpha_r\geq{\sigma\frac{\sqrt{\zeta_j}}{\sqrt{3}}}\mathcal{R}\{d_j\}, d_j=\frac{3+i}{\sqrt{2}},\frac{3-i}{\sqrt{2}}\\
\alpha_r\leq{\sigma\frac{\sqrt{\zeta_j}}{\sqrt{3}}}\mathcal{R}\{d_j\}, d_j=\frac{-3+i}{\sqrt{2}},\frac{-3-i}{\sqrt{2}}\end{cases}
\end{eqnarray}

 \begin{eqnarray}\nonumber
\mathcal{C}_2=\begin{cases}
\alpha_i\geq {\sigma\sqrt{\frac{\zeta_j}{3}}}\mathcal{I}\{d_j\}, d_j=\frac{\pm 1+i}{\sqrt{2}},\frac{\pm 3+i}{\sqrt{2}},\\
\alpha_i\leq{\sigma\sqrt{\frac{\zeta_j}{3}}}\mathcal{I}\{d_j\}, d_j=\frac{\pm 1-i}{\sqrt{2}},\frac{\pm 3-i}{\sqrt{2}}\end{cases}
\end{eqnarray}
 For the 16-QAM modulation, the constraints $\mathcal{C}_1$, $\mathcal{C}_2$ can be expressed as
 \begin{eqnarray}\nonumber
\mathcal{C}_1=\begin{cases}\alpha_r=\sigma\sqrt{\frac{\zeta_j}{5}}\mathcal{R}\{d_j\}, d_j=\frac{\pm 1+\pm i}{\sqrt{2}}, \frac{\pm 1+\pm 3i}{\sqrt{2}}\\
\alpha_r\geq{\sigma\sqrt{\frac{\zeta_j}{5}}}\mathcal{R}\{d_j\}, d_j=\frac{3+i}{\sqrt{2}},\frac{3-i}{\sqrt{2}},\frac{3+3i}{\sqrt{2}},\frac{3-3i}{\sqrt{2}}\\
\alpha_r\leq{2\sigma\sqrt{\frac{\zeta_j}{5}}}\mathcal{R}\{d_j\}, d_j=\frac{-3+i}{\sqrt{2}},\frac{-3-i}{\sqrt{2}}, \frac{-3+3i}{\sqrt{2}}, \frac{-3-3i}{\sqrt{2}}\end{cases}
\end{eqnarray}

 \begin{eqnarray}\nonumber
\mathcal{C}_2=\begin{cases}\alpha_i=\sigma\sqrt{\frac{\zeta_j}{5}}\mathcal{I}\{d_j\}, d_j=\frac{\pm 1+\pm i}{\sqrt{2}}, \frac{\pm 3+\pm i}{\sqrt{2}},\\ \alpha_i\geq \sigma\sqrt{\frac{\zeta_j}{5}}\mathcal{I}\{d_j\}, d_j=\frac{\pm 1+3i}{\sqrt{2}}, \frac{\pm 3+3i}{\sqrt{2}}\\\alpha_i\leq{\sigma\sqrt{\frac{\zeta_j}{5}}}\mathcal{I}\{d_j\}, d_j=\frac{\pm 1-3i}{\sqrt{2}},\frac{\pm 3-3i}{\sqrt{2}}\end{cases}
\end{eqnarray}
\begin{figure}[t]
\hspace{-0.7cm}\includegraphics[scale=0.65]{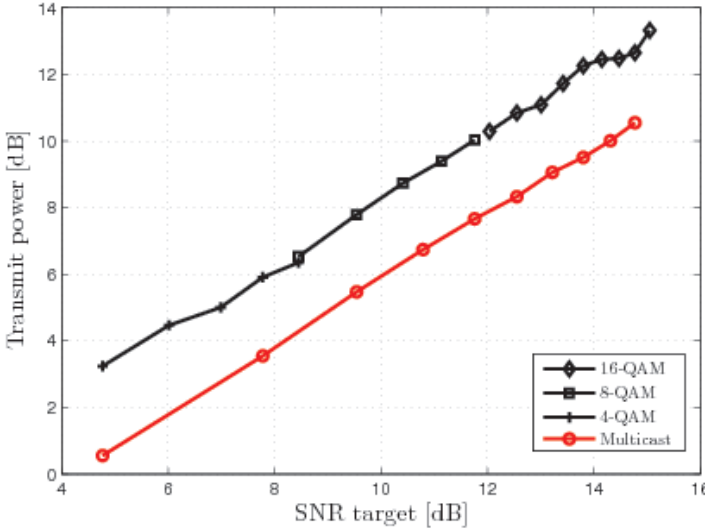}
\caption{\label{ci1}\textit{\label{fig1}\small Transmit power vs. the target SNR, $M=2$,$K=2$, $\sigma^2=0dB$, $\gamma_{\circ}=0dB$ $\zeta_j=\zeta_{th},\forall j\in K$.}}
\end{figure}

Fig. (\ref{ci1}) depicts the amount of the required  transmit power $\|\mathbf{x}\|^2$ to achieve certain target SNR exploiting symbol-level precoding CIPM. It can be noted that the PHY-multicasting presents a lower-bound for the proposed technique. It can be noted that the performance of different modulations a continuous pattern with increasing the modulation order. Moreover, the power consumption increases linearly in dB with increasing the SNR target.
\begin{figure}[t]
\hspace{-0.5cm}\includegraphics[scale=0.65]{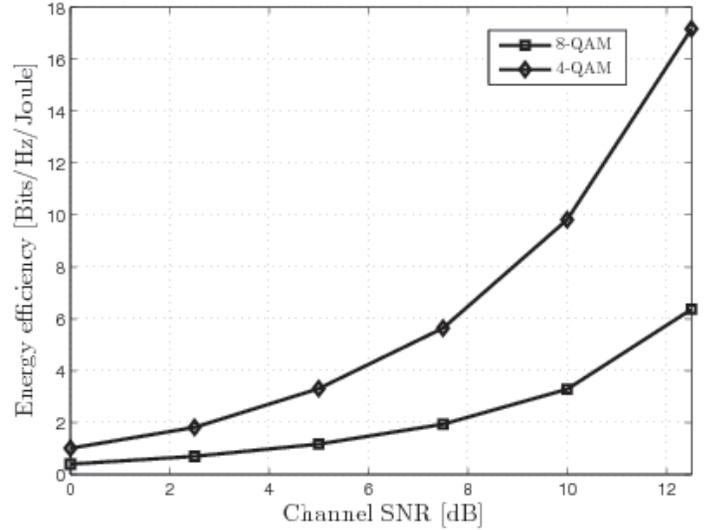}
\caption{\label{ci2}\textit{\label{fig2}\small Energy efficiency $\eta$ vs. the channel SNR $\sigma^2$, $M=2$,$K=2$, $\sigma^2=0dB$,  $\zeta_j=\zeta_{th},\forall j\in K$, $\zeta_{th}$ for 8-QAM equals to 9 dB and $\zeta_{th}$ for 4-QAM equals to 6 dB. }}
\end{figure}

Fig. (\ref{ci2})  depicts the comparison between the energy efficiency of 8-QAM and 4-QAM respectively. In this figure, we assume that the SNR targets for 8-QAM and 4-QAM equal to $9$ dB $6$ dB respectively to fit the requirement of having higher SNR targets. Although 4-QAM has lower rate with increasing the channel SNR, It can be noted that it has higher energy efficiency. This can be explained by the fact that SER in 8-QAM is higher which makes the numerator in the energy efficiency more sensitive to the SER. Moreover, the power consumption in 8-QAM is higher due to higher SNR requirement, which results in higher energy efficiency.

\begin{figure}[t]
\hspace{-0.1cm}\includegraphics[scale=0.65]{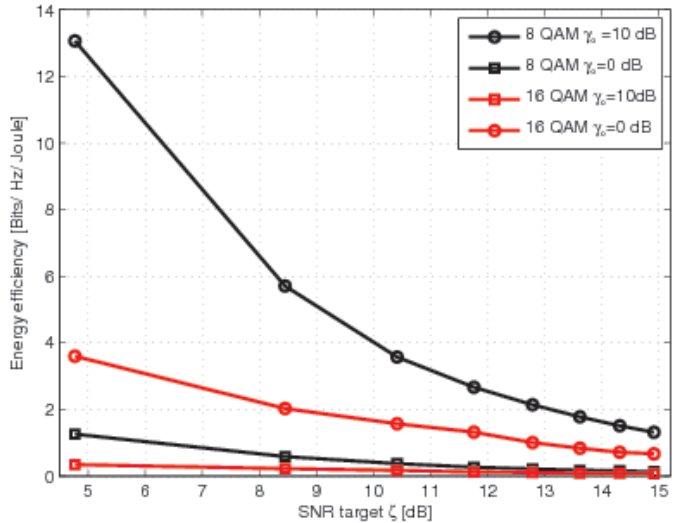}
\caption{\label{ci3}\textit{\label{fig3}\small Energy efficiency $\eta$ vs. the target SNR, $M=3$,$K=2$, $\zeta_j=\zeta_{th},\forall j\in K$, $\sigma^2=0dB$.}}
\end{figure}

Fig. (\ref{ci3}) depicts the energy efficiency performance of 16-QAM and 8-QAM with respect to SNR target $\zeta_{th}$.  It can be noted that the energy efficiency decreases with increasing SNR target $\zeta_{th}$, we assume that 8-QAM and 16-QAM have the same $\zeta_{th}$ to see the impact of SER. It can be noted that constructive interference for 8-QAM has higher energy efficiency due to lower SER in comparison to 16-QAM. 

\section{conclusions}
In this paper, we utilized jointly CSI and DI in symbol based precoding to exploit received interfering
signal as useful energy in constructive interference precoding. In these cases, the precoding design exploits the
overlap in users' subspace instead of mitigating it. This fact enabled us
to find the connection between the constructive interference precoding and
multicast precoding wherein no interference should be mitigated. In this work, we propose precoding techniques that extends the concept of constructive interference to multi-level constellation. Therefore,
we found the solution for power minimization considering two inputs scenario:
the optimal input and the constrained constellation. From their closed formulations,
we concluded that their transmissions should span the subspaces of each user. From the numerical results, it can be concluded that the energy efficiency is higher for lower modulation order.

\end{document}